\documentclass{article}
\makeatletter
\@addtoreset{equation}{section}

\makeatother

\begin{document}

\begin{flushright}
Oct 2005

KUNS-1991
\end{flushright}

\begin{center}

\vspace{5cm}

{\LARGE 
\begin{center}
Closed String Tachyon Condensation 

in Supercritical Strings and RG Flows
\end{center}
}

\vspace{2cm}

Takao Suyama \footnote{e-mail address : suyama@gauge.scphys.kyoto-u.ac.jp}

\vspace{1cm}

{\it Department of Physics, Kyoto University,}

{\it Kitashirakawa, Kyoto 606-8502, Japan }

\vspace{4cm}

{\bf Abstract} 

\end{center}

We show an explicit relation between an RG flow of a two-dimensional gravity and an on-shell 
tachyon condensation in the corresponding string theory, in the case when the string theory is 
supercritical. 
The shape of the tachyon potential in this case can, in principle, be obtained by examining various 
RG flows. 
We also argue that the shape of tachyon potential for a (sub)critical string can be obtained by 
analyzing a supercritical string which is obtained from the (sub)critical string.

\newpage

\vspace{1cm}

\section{Introduction}

\vspace{5mm}

Understanding closed string tachyons and their condensation is still a challenging problem in 
string theory. 
Basically, people seem to believe that an analogous phenomena with the case of open string tachyons 
would occur even for the case of closed string tachyons. 
That is, a perturbative vacuum, which has closed string tachyons, of a string theory is an unstable 
``state'' in a string theory with a stable vacuum, for example, a superstring. 
The stabilization of the tachyonic vacuum would be realized by a condensation of the closed string 
tachyons. 
There were preliminary works on this subject \cite{Halpern}. 

This line of thought is acceptable with ease when the tachyons are localized around some region of the 
target space at which the environment is different from other regions. 
One such example which has been studied intensively is string theory on 
non-supersymmetric orbifolds \cite{APS}. 
In such a case, tachyons are localized around the orbifold singularity, so one can regard that the 
singularity is an unstable object. 
Then it is natural to expect that a condensation of the tachyons would get rid of the singularity 
in order to stabilize the theory. 
It has been checked in different ways \cite{APS}\cite{Vafa}\cite{HKMM} 
that this kind of mechanism actually exists. 
However, the techniques used in the analyses have a common 
limitation; they can apply only to cases in which 
a perturbation of the worldsheet theory by a tachyon vertex operator 
does not change the central charge. 
It has been believed that an RG flow induced by a tachyon vertex operator 
can be a guide to find the final 
state of a condensation of the tachyon. 
Therefore, the final state of the condensation can be described by a string theory only if the 
central charge is kept fixed during the condensation. 
As a result, closed string tachyons localized in a non-compact space has been understood 
comparatively, but the other tachyons, namely, tachyons localized in a compact space and bulk tachyons, 
remain to be understood. 
Recently, there appear some papers on closed string tachyon condensations 
\cite{She}\cite{BH}\cite{IKS}\cite{Narayan}\cite{FHL}\cite{HT}. 

In our previous paper \cite{Suyama}, we argued closed string tachyons localized in compact spaces. 
We related on-shell tachyon condensations to RG flows of two-dimensional gravity theories 
(two-dimensional field theories coupled to gravity), and proposed a final state of a condensation for 
some explicit examples of tachyonic string theory. 
We argued a condensation starting from a critical string (string without dilaton background), 
but in this case 
there is a difficulty in determining which is the time direction, since the RG flow is in fact 
related to an evolution along the Euclideanized time. 
Although it would be reasonable to expect that a proper Wick rotation enables one to extract some 
information on the condensation, it is not yet clear how it can be realized. 

In this paper, we consider a slightly different situation which is easier to analyze 
than the previous cases. 
We discuss a condensation starting from a supercritical string (string with a timelike dilaton 
gradient). 
In this case, RG flows can be naturally related to evolutions in string theory along the ordinary 
time, and therefore the correspondence of the RG flows to time evolutions can be shown explicitly. 
We also show that this analysis can be used to understand a condensation starting from critical 
string, by relating the critical string to a supercritical string which shares the same information 
on the tachyon potential with the critical string. 

This paper is organized as follows. 
In section \ref{2}, we review a relation between a two-dimensional gravity and a string theory. 
In section \ref{3}, an RG flow in a two-dimensional gravity is related to a time evolution, 
in particular an on-shell tachyon condensation. 
In section \ref{4}, 
an effective action of graviton, dilaton and tachyon is argued, and the correspondence 
between an RG flow and a time-dependent solution of the effective action is explicitly shown. 
In section \ref{5}, 
the tachyon potential in critical strings (and sub-critical strings, that is, strings 
with spacelike dilaton gradients) is discussed. 
Appendix \ref{derivation} summarizes a derivation of the induced action of the Liouville field.

\vspace{1cm}

\section{Two-dimensional gravity as a string theory} \label{2}

\vspace{5mm}

In this section, we review a relation between a worldsheet RG flow in a two-dimensional gravity and 
a time evolution of a string theory \cite{Suyama}, based on \cite{D}\cite{DK}. 

Consider a two-dimensional field theory, not necessarily conformal, 
specified by the action $S(X;g)$, where $X$ collectively 
denotes matter fields and $g$ is a metric on the worldsheet $\Sigma$. 
Its partition function is defined by 
\begin{equation}
Z = \int \frac{{\cal D}X{\cal D}g}{\mbox{vol(diff)}}\ e^{-S(X;g)-\mu_0\int d^2\sigma\sqrt{g}}.
   \label{Z}
\end{equation}
We introduced the bare cosmological constant $\mu_0$ whose purpose will be explained below. 

One can always fix the diffeomorphism invariance by choosing a gauge slice 
\begin{equation}
g = e^{\varphi}\hat{g}(\tau), 
   \label{fixing}
\end{equation}
where $\hat{g}(\tau)$ is a representative metric of an equivalence class of metrics parametrized 
by a finite number of moduli parameters collectively denoted by $\tau$. 
The partition function (\ref{Z}) becomes, after gauge fixing, 
\begin{equation}
Z = \int {\cal D}X\ d\tau{\cal D}b{\cal D}c{\cal D}\varphi\ 
    e^{-S(X;e^\varphi\hat{g},\mu_0)-S(b,c;\hat{g})}, 
\end{equation}
where the cosmological constant term is included in the matter action. 
Note that the ghost action is Weyl invariant. 

The path integral measures ${\cal D}X$ etc. are defined by using $g$, and therefore they depend on 
$\varphi$ as well as $\hat{g}$. 
We assume that the $\varphi$-dependence of the measures can be extracted to provide an exponentiated 
local action, that is, 
\begin{equation}
Z = \int {\cal D}_{\hat{g}}X\ d\tau{\cal D}_{\hat{g}}b{\cal D}_{\hat{g}}c{\cal D}_{\hat{g}}\varphi\ 
    e^{-S(X,\varphi;\hat{g},\mu_0)-S(b,c;\hat{g})}. 
      \label{Z1}
\end{equation}

The gauge fixing condition (\ref{fixing}) is not the only choice. 
For example, one can also choose 
\begin{equation}
g = e^{\varphi'}\cdot e^\omega\hat{g}(\tau), 
\end{equation}
for an arbitrary function $\omega$. 
For this gauge fixing, one obtains 
\begin{equation}
Z = \int {\cal D}_{e^\omega\hat{g}}X\ d\tau{\cal D}_{e^\omega\hat{g}}b{\cal D}_{e^\omega\hat{g}}c
    {\cal D}_{e^\omega\hat{g}}\varphi'\ e^{-S(X,\varphi';e^\omega\hat{g},\mu_0)
    -S(b,c;e^\omega\hat{g})}. 
       \label{Z2}
\end{equation}
The latter partition function (\ref{Z2}) is obtained from (\ref{Z1}) by a Weyl transformation and 
a rename of the field $\varphi$. 
The equivalence of (\ref{Z1}) and (\ref{Z2}) implies that the gauge-fixed theory (\ref{Z1}) 
is a CFT with vanishing total central charge. 
Therefore, one can regard the partition function (\ref{Z1}) as a perturbative contribution to a 
partition function of 
a string theory, in which $\varphi$ is also regarded as a coordinate field of a string. 

In general, it is difficult to determine the action $S(X,\varphi;\hat{g},\mu_0)$ from the original 
matter action $S(X;g)$. 
However, one can obtain the action explicitly when the original matter action is conformal. 
The resulting action is 
\begin{equation}
S(X,\varphi;\hat{g},\mu_0) 
= S(X;\hat{g})+S(b,c;\hat{g})
+\frac{25-c_m}{48\pi}\int d^2\sigma\sqrt{\hat{g}}
\left[ \frac12(\hat{\nabla}\varphi)^2+\hat{R}\varphi \right],  
   \label{GFaction}
\end{equation}
where $c_m$ is the central charge of the matter CFT. 
A derivation of this action is reviewed in appendix \ref{derivation}. 
Note that, in general, there is the Liouville potential term $\mu e^{b\varphi}$ in (\ref{GFaction}). 
We chose the bare cosmological constant $\mu_0$ so that $\mu=0$, which is usually assumed to be 
possible. 

Interestingly, if the matter CFT has a central charge $c_m>25$, then the kinetic term of $\varphi$ has 
the wrong sign, implying that $\varphi$ describes the {\it time direction} of a target spacetime of 
the corresponding string theory\footnote{In $c_m<25$ case, 
$\varphi$ is regarded as the Euclideanized time direction.}. 
This is one of the key observation for relating a worldsheet RG flow and a time evolution in a string 
theory. 

It is convenient to rescale $\varphi$ so that the action of $\varphi$ has the usual coefficients. 
Then one obtains 
\begin{equation}
\frac1{4\pi}\int d^2\sigma\sqrt{\hat{g}}\left[ -(\hat{\nabla}\varphi)^2
-\sqrt{\frac{c_m-25}{6}}\hat{R}\varphi \right]. 
   \label{Liouville}
\end{equation}
Now one can easily calculate the total central charge of the CFT as follows, 
\begin{equation}
c_m-26+(1+(25-c_m)) = 0, 
\end{equation}
which was expected from the general argument.

\vspace{1cm}

\section{Worldsheet RG flow} \label{3}

\vspace{5mm}

\subsection{General argument}

\vspace{5mm}

We would like to consider an RG flow of a two-dimensional gravity. 
What is meant by the ``RG flow of a two-dimensional gravity'' is as follows. 
Consider a generic two-dimensional field theory coupled to gravity. 
In general, the theory may be deformed as one changes the scale of the theory. 
The scale can be chosen, for example, by introducing a desired value of the cut-off for the momentum, 
\begin{equation}
g_{\alpha\beta}p^\alpha p^\beta \le \Lambda^2, 
\end{equation}
and integrating out the large momentum part of fields in the theory. 
The change of the theory as $\Lambda$ varies from $+\infty$ to $0$ defines a one-parameter family of 
two-dimensional gravity theories. 
Then one can define a flow in the space of two-dimensional gravity theories, called the RG flow. 
However, if $g$ is not just a background and is integrated over all possible metrics, then the change 
of $\Lambda$ is absorbed by a rescaling of $g$, and the theory is kept intact. 
Therefore, one also has to fix the scale of $g$ so that the change of scale makes some non-trivial 
effects. 
This fixing can be done, for example, by fixing the area of the worldsheet, 
\begin{equation}
\int d^2\sigma\sqrt{g} = A. 
   \label{area}
\end{equation}
Now one can investigate the RG flow by varying $\Lambda$ with a fixed $A$, but equivalently, one can 
see the same flow by varying $A$ with a fixed $\Lambda$. 
In fact, one can choose the fixed value of 
$\Lambda$ as large as possible, so that the momentum cut-off $\Lambda$ can be 
effectively eliminated. 

By the gauge choice (\ref{fixing}), the condition (\ref{area}) becomes 
\begin{equation}
\int d^2\sigma\sqrt{\hat{g}}\ e^\varphi = A. 
\end{equation}
Therefore, the change of $A$ corresponds to a shift of the zero mode $\varphi_0$ of $\varphi$. 
By combining this relation with the relation between $\Lambda$ and $A$, one finds that 
$\varphi_0\to -\infty\ (+\infty)$ corresponds to the UV (IR) limit of the RG flow. 

Recall that $\varphi$ is the time direction if the matter theory is conformal. 
This implies that, at least around the UV and the IR limit of the RG flow, the flow would describe 
a time evolution of the corresponding string theory, provided that the central charge of the IR limit 
is greater than $25$. 
And, as we have shown, there is a CFT (\ref{Z1}) 
with vanishing total central charge which corresponds to 
a general two-dimensional gravity. 
Therefore, it is natural to expect that the whole RG flow of the two-dimensional gravity describes 
the whole process of a time evolution of the corresponding string theory. 

\vspace{5mm}

\subsection{Small perturbation}

\vspace{5mm}

As mentioned above, it is difficult, in general, to obtain a CFT (\ref{Z1}) from a two-dimensional 
gravity which one would like to study. 
By the same reason, it is difficult to obtain a two-dimensional gravity whose RG flow corresponds to 
a tachyon condensation which one would like to study. 
However, if one is only interested in the initial and the final stage of the condensation, it would be 
rather easy to find which RG flow one has to consider. 

For example, suppose that we would like to study 
a string theory, say bosonic string compactified on a manifold 
$M$ with dim$M=d$. 
Consider a CFT whose action is 
\begin{equation}
S = \frac1{4\pi}\int d^2\sigma\sqrt{g}\nabla X^m\cdot\nabla X^n\delta_{mn}
    +S_M(g), 
       \label{matterCFT}
\end{equation}
where $S_M$ is a CFT describing a string living in $M$, and $m,n=1,\cdots,25-d$. 
We have set $\alpha'=1$. 
The central charge of this CFT is $c_m=25$. 
Therefore, the Liouville field $\varphi$ appearing 
after the gauge fixing does not have a linear dilaton background\footnote{Since the overall 
coefficient of the Liouville action vanishes in this case, some 
regularization, say an infinitesimal shift of the central charge which will be eliminated after 
rescaling $\varphi$, would be necessary.} , and the gauge fixed theory 
(\ref{Z1}) is the bosonic string which we would like to study. 

Suppose that the bosonic string has a tachyon in the mass spectrum. 
Let $V(k)$ be the vertex operator of the tachyon, where $k$ is its momentum in the flat directions. 
Then $V=V(k=0)$ may have the weight $\Delta=\bar{\Delta}<1$, that is, this is a relevant operator. 
One can consider a perturbation of (\ref{matterCFT}) by adding the term 
\begin{equation}
\lambda\int d^2\sigma\sqrt{g}\ V,
   \label{perturb}
\end{equation}
to (\ref{matterCFT}). 
This perturbation induces a non-trivial RG flow. 
As long as the added term (\ref{perturb}) can be treated as a small perturbation, the gauge fixed 
theory (\ref{Z1}) can be obtained explicitly as follows, 
\begin{equation}
S = \frac1{4\pi}\int d^2\sigma\sqrt{\hat{g}}\nabla X^\mu\cdot\nabla X^\nu\eta_{\mu\nu}
    +S_M(\hat{g})+\lambda\int d^2\sigma\sqrt{\hat{g}}\ e^{\alpha X^0}V, 
       \label{fixedCFT}
\end{equation}
where $\mu,\nu=0,\cdots,25-d$ and $X^0=\varphi$. 
The exponent $\alpha$ is determined such that the operator $e^{\alpha X^0}V$ is marginal with 
respect to the unperturbed theory. 
Note that the same procedure can be applied to $c_m\ne25$ cases, in which a linear dilaton background 
(\ref{Liouville}) appears, 
and also to the cases in which $V$ is 
irrelevant. 
In the following, we will consider $c_m>25$ cases. 

The condition for the operator $e^{\alpha X^0}V$ to be marginal is 
\begin{equation}
\Delta + \frac14\alpha(\alpha+Q) = 1, 
\end{equation}
where $Q=2\sqrt{\frac{c_m-25}6}$. 
Then one obtains 
\begin{eqnarray}
\alpha_\pm
&=& \frac12\Bigl(-Q\pm\sqrt{Q^2-16(\Delta-1)}\Bigr) \nonumber \\
&=& -\sqrt{\frac{c_m-25}6}\pm\sqrt{\frac{c_m-1}6-4\Delta}. 
       \label{exponent}
\end{eqnarray}
In particular, $\alpha_\pm$ is complex if 
\begin{equation}
\Delta>\frac{c_m-1}{24}>1, 
\end{equation}
and  
\begin{equation}
\alpha_+>0>\alpha-,
\end{equation}
only for $\Delta<1$. 
If Re$(\alpha_+)>0$, then the term $e^{\alpha X^0}V$ grows as $X^0$ becomes large, implying that 
the RG flow corresponds to a homogeneous condensation of $V$. 
Therefore, one can find the RG flow of the two-dimensional gravity corresponding to a tachyon 
condensation, by taking as $V$ the vertex operator of the tachyon. 

If the action (\ref{fixedCFT}) indeed describes an on-shell process, the operator 
$e^{\alpha_\pm X^0}V$ 
describes an on-shell tachyon background. 
Therefore, there should exist a classical solution 
\begin{equation}
g_{\mu\nu}=\eta_{\mu\nu}, \hspace{5mm} 
\Phi(t) = -2\sqrt{\frac{c_m-25}6}t, \hspace{5mm}
T(t) \propto e^{\alpha_\pm t}, 
   \label{RGsoln}
\end{equation}
of the effective action, as long as $|T(t)|$ is small. 
Moreover, the flat metric should be a solution even when $|T|$ is not small. 
This is because the part of the flat spatial directions in (\ref{matterCFT}) is decoupled from another 
part $S_M(g)$ which is deformed by the perturbation (\ref{perturb}). 
Since we identified the time coordinate $t$ with the Liouville field $\varphi$ (or $X^0$), 
there is no ambiguity in the reparametrization of the time coordinate.

\vspace{1cm}

\section{Classical solutions for rolling tachyon} \label{4}

\vspace{5mm}

\subsection{Asymptotic solutions}

\vspace{5mm}

In order to check the correspondence of the RG flow to an on-shell condensation, 
it is necessary to obtain an effective action of the string theory which describes the above on-shell 
tachyon condensation. 
It is in general difficult since there is no convincing reason to exclude from the effective action 
all massive states in string theory, while keeping a tachyon with $m^2$ of order of the string 
scale. 
However, the situation would be better when one restrict oneself to consider only the initial or the 
final stage of the process. 
Suppose that the process would start from or end at a low energy background configuration. 
Then in the beginning or the final stage of the process, every fields vary slowly and no massive 
states would be excited unless there already exists a non-trivial background for massive states. 
Therefore, it would be reasonable to exclude all massive fields from the effective action as long as 
one only consider an asymptotic behavior of classical solutions. 

In the following, we analyze the effective action of a tachyon $T$, the graviton $g_{\mu\nu}$ and the 
dilaton $\Phi$. 
The form of the effective action whose terms include up to two derivatives is 
\begin{equation}
S = \frac1{2\kappa^2}\int d^Dx\sqrt{-g}\ e^{-2\Phi}\Bigl[ 
    f_1(T)R+4f_2(T)(\nabla\Phi)^2+2f_3(T)\nabla\Phi\cdot\nabla T-f_4(T)(\nabla T)^2-2V(T) \Bigr]. 
      \label{general}
\end{equation}
The functions $f_i(T)$ $(i=1,2,3,4)$ cannot be determined only by symmetries. 
Since this action should reduce to the well-known effective action when $T=0$, one obtains 
$f_1(0) = f_2(0) = 1$, and $f_3(0)=0$ since there should be no mixing between $T$ and $\Phi$. 
One can always choose $f_4(T)=1$ by a suitable field redefinition of $T$. 
We consider processes in which $|T|<<1$. 
Under this condition, one can ignore the $T$-dependence of $f_i(T)$. 
That is, we analyze 
\begin{equation}
S = \frac1{2\kappa^2}\int d^Dx\sqrt{-g}\ e^{-2\Phi}\Bigl[ 
    R+4(\nabla\Phi)^2-(\nabla T)^2-2V(T) \Bigr]. 
      \label{simpleaction}
\end{equation}
This effective action is the one studied recently in \cite{YZ} in which some properties shown below 
are already mentioned\footnote{The similar analysis was also done in \cite{PASCOS}.}. 
The solution was analyzed numerically in \cite{DGKK}\cite{Suyama2}.  
In addition, one can use the following simple form of the potential,   
\begin{equation}
V(T) = V_0+\frac12m^2T^2.
\end{equation}
Since the linear dilaton background (\ref{Liouville}) in the flat metric with $T=0$ should be a 
solution of this effective action, one obtains 
\begin{equation}
V_0 = \frac{c_m-25}3. 
   \label{V_0}
\end{equation}
We analyze a homogeneous field configuration, by making an ansatz, 
\begin{equation}
ds^2 = -dt^2+a(t)^2\delta_{mn}dx^mdx^n, \hspace{5mm} 
T=T(t), \hspace{5mm} \Phi=\Phi(t). 
   \label{ansatz}
\end{equation}
The equations of motion reduces to the following ones, 
\begin{eqnarray}
\frac{D-2}{2(D-1)}h^2 &=& \frac12\dot{T}^2+V(T)-2\phi^2+2h\phi, 
   \label{a} \\
\dot{\phi} &=& -h\phi+2\phi^2-V(T), 
   \label{phi} \\
\ddot{T} &=& -h\dot{T}+2\phi\dot{T}-V'(T), 
   \label{T}
\end{eqnarray}
where 
\begin{equation}
h=(D-1)\frac{\dot{a}}a, \hspace{5mm} \phi=\dot{\Phi}. 
\end{equation}
One can show that $h=0$ is consistent with these equations of motion, and the equations to be solved 
are then 
\begin{eqnarray}
2\phi^2 &=& \frac12\dot{T}^2+V(T), 
   \label{eq}\\
\dot{\phi} &=& 2\phi^2-V(T). 
\end{eqnarray}
In fact, these two equations imply (\ref{T}) with $h=0$. 
Note that this is also the case for general $V(T)$. 
The fact that the flat string metric is a solution is consistent with the relation between the RG 
flows and the on-shell processes, as mentioned before. 

The $h=0$ solution is stable in the following sense. 
From the equations of motion, one can derive 
\begin{equation}
\Bigl[ \frac{D-2}{D-1}h-2\phi \Bigr]\Bigl[ \dot{h}-h(2\phi-h) \Bigr] = 0. 
\end{equation}
Under the condition $\frac{D-2}{D-1}h=2\phi$, (\ref{a}) does not have a non-trivial solution, and 
one obtains 
\begin{equation}
\dot{h} = h(2\phi-h). 
   \label{stability}
\end{equation}
Since $\phi<0$ around any fixed point when $c_m>25$, as one can see from (\ref{Liouville}), 
this equation implies that $h=0$ is a stable 
fixed point. 

When $h=0$, one can show that 
\begin{equation}
\dot{\phi} = \frac12\dot{T}^2. 
\end{equation}
This indicates that $\dot{\Phi}$ does not decrease. 
Since $\phi<0$ as mentioned above, and due to (\ref{eq}), $\phi$ has the upper bound $-\sqrt{2V_0}$ 
at which the tachyon $T$ stops at $T=0$ when $m^2>0$, that is, $T=0$ is a minimum of $V(T)$. 
In other words, the tachyon always stops at a minimum of the potential provided that the 
value of the potential minimum is positive. 
The monotonicity of $\dot{\Phi}$ would be related to the monotonicity of the c-function in the 
corresponding RG flow \cite{c-th}. 
Recall that the central charge for the Liouville field $\varphi$ is 
$c_L=1-\frac32Q^2$ in our convention, 
and $Q$ is proportional to the gradient of the dilaton. 
As we chose a negative value for $\dot{\Phi}$, then $\dot{\Phi}^2$ decreases and $c_L$ 
increasing, compensating the decrease of the matter central charge along the RG. 
In this way, the total central charge is kept fixed to be zero, a necessary condition to interpret the 
RG flow as an on-shell process of a string theory. 

When $\phi<0$, (\ref{T}) tells that $T$ feels a friction, and as a result, $T$ can stop at a 
minimum of the potential $V(T)$. 
This final field configuration would be related to the IR fixed point of the RG flow. 
Note that $\phi<0$ ensures that one can keep the string coupling as weak as possible along the 
process (in fact, the string coupling decreases in time), 
so quantum corrections to the classical solution is negligible. 

\vspace{5mm}

Let us examine the time-dependence of the solution around the trivial solution $T=0,\ 2\phi^2=V_0$. 
Assume that $T(t),\ \dot{T}(t)$ are both small. 
Then, at the linearized level, one obtains 
\begin{eqnarray}
\dot{\phi} &=& 0, \\
\ddot{T} &=& -\sqrt{2V_0}\dot{T}-m^2T, 
\end{eqnarray}
from the equations of motion. 
Note that, as show above, $h=0$ is a solution even beyond the linearized level. 
These equations are easily solved, and one obtains 
\begin{equation}
\Phi(t) = \Phi_0-\sqrt{2V_0}t, \hspace{5mm} 
T(t) = Ae^{a_+t}+Be^{a_-t}, 
   \label{Gsoln}
\end{equation}
where 
\begin{equation}
a_\pm = -\sqrt{\frac{V_0}2}\pm\sqrt{\frac{V_0}2-m^2}. 
   \label{a+-}
\end{equation}
This solution coincides with (\ref{RGsoln}) when $V_0$ is given by (\ref{V_0}) and 
\begin{equation}
\frac{m^2}4 = \Delta-1, 
\end{equation}
as expected\footnote{For the light-cone quantization of string theory in the linear dilaton background, 
see \cite{quantization}.}. 
This coincidence indicates 
nothing but the fact that the equation of motion of the tachyon is also obtained 
by requiring the conformal invariance of the worldsheet theory. 

When $m^2<0$, that is $T=0$ is a maximum of $V(T)$, $e^{a_+t}$ is always growing, which should be the 
desired behavior for a perturbation by a tachyon vertex ($\Delta<1$). 
On the other hand, when $m^2>0$, a minimum of $V(T)$, $T(t)$ is always damping exponentially, in good 
correspondence to an irrelevant perturbation. 
For $0<m^2<\frac{V_0}2$ it is overdampling, and for $m^2>\frac{V_0}2$ it exhibits a damped 
oscillation. 

\vspace{5mm}

It has been shown that the RG flows in the vicinity of UV and IR fixed points are well described by 
classical solutions around a maximum and a minimum of the potential $V(T)$, respectively. 
In other words, the physical picture 
gained in the study of open string tachyon condensation still persists in the case of closed 
string tachyon condensation localized in a compact space, provided that $c_m>25$. 
When there is an RG flow which connects a UV CFT with a tachyon and an IR CFT, then in the effective 
action there is a potential which has a maximum and a minimum which are next to each 
other. 
The difference of the height of these extrema is 
\begin{equation}
V(\mbox{UV})-V(\mbox{IR}) = \frac{c_{m,UV}-c_{m,IR}}3. 
   \label{height}
\end{equation}
Therefore, the shape of the tachyon potential can be determined, except for the positions 
of the extrema, by examining various RG flows. 
It will be very exciting if the relation (\ref{height}) 
is confirmed, for example, by using string field theory.

\vspace{5mm}

\subsection{Beyond linearized approximation}

\vspace{5mm}

Let us argue the form of the effective action beyond the linearized approximation. 
One would like to find a solution of (\ref{general}), but the action has many unknown functions 
$f_i(T)$, 
and it is difficult to extract general properties of solutions. 
Fortunately, some of the functions can be fixed from the following argument on a worldsheet theory. 

Consider a worldsheet theory in a background metric, dilaton and tachyon, 
\begin{equation}
S 
= \frac1{4\pi}\int d^2\sigma\sqrt{h} \Bigl[ \nabla X^\mu\nabla X^\nu g_{\mu\nu}(X) 
  +\Phi(X)R_h \Bigr]+S_M(h)+\int d^2\sigma\sqrt{h}\ T(X)V, 
\end{equation}
where $\nabla$ and $R_h$ are the covariant derivative and the scalar curvature, respectively, 
defined by the 
worldsheet metric denoted by $h$ in this subsection. 
Note that $h$ is not dynamical. 

Suppose that $T(X)=T=$ const. and $g_{\mu\nu},\ \Phi$ does not depend on fields of $S_M(h)$. 
Then the worldsheet theory consists of two theories $S(X;g,\Phi), S_M(T)$ decoupled from each other. 
The energy-momentum tensor of the total theory is a sum of those of two theories, and therefore, 
the beta-functionals are also a sum of two functionals. 
For examples, 
\begin{eqnarray}
\beta_{\mu\nu}(g) &=& R_{\mu\nu}+2\nabla_\mu\nabla_\nu\Phi+\cdots, \\
\beta(\Phi) &=& \frac{D-26}6-\frac12\nabla^2\Phi+(\nabla\Phi)^2+\cdots, 
\end{eqnarray}
where $\cdots$ indicates terms including $T$, and $\mu,\nu=0,\cdots,D-1$. 
These beta-functionals should be derived from (\ref{general}) as the equations of motion, even when $T$ 
is not small. 
Therefore, one concludes that $f_1(T)=f_2(T)=1$. 
One may redefine $T$ so that $f_4(T)=1$. 
Now the action to be analyzed becomes slightly simple as follows, 
\begin{equation}
S = \frac1{2\kappa^2}\int d^Dx\sqrt{-g}\ e^{-2\Phi}\Bigl[ 
    R+4(\nabla\Phi)^2+2f(T)\nabla\Phi\cdot\nabla T-(\nabla T)^2-2V(T) \Bigr]. 
      \label{restricted}
\end{equation}
As mentioned before, $f(T)$ should satisfies $f(0)=0$, but otherwise one cannot impose further 
constraints on the form of $f(T)$ by the above observation. 
One way to restrict $f(T)$ would be to require that solutions of the equations of motion derived from 
(\ref{restricted}) have a dilaton solution whose derivative grows monotonically.

\vspace{1cm}

\section{Lift of critical strings to supercritical strings} \label{5}

\vspace{5mm}

As we have shown, an RG flow of a two-dimensional gravity with $c_m>25$ describes an on-shell 
tachyon condensation from a potential maximum to a minimum. 
By using this relation, one can qualitatively know the shape of the tachyon potential from various 
RG flows. 
However, for the $c_m\le25$ cases, RG flows cannot be directly related to a time evolution since 
in this case the Liouville field describes a spacelike direction, as one can see from 
(\ref{GFaction}). 
In addition, classical solution of the effective action (\ref{simpleaction}), in which $V(T)$ has a 
negative value at a minimum, does not always converge to a static solution corresponding to the 
potential minimum \cite{YZ}. 
Therefore, the relation between RG flows and time evolutions is not yet unclear in this case, 
although they 
might be related by a suitable Wick rotation. 

In this section, we argue that at least the shape of the potential $V(T)$ can be read off from 
RG flows, even in the $c_m\le25$ cases. 
The idea is that one can relate to a (sub)critical string $(c_m\le25)$ a 
supercritical string $(c_m>25)$, 
without changing the shape of the potential. 

Consider, again, the following worldsheet theory, 
\begin{equation}
S 
= \frac1{4\pi}\int d^2\sigma\sqrt{h} \Bigl[ \nabla X^\mu\nabla X^\nu g_{\mu\nu}(X) 
  +\Phi(X)R_h \Bigr]+S_M(h)+\int d^2\sigma\sqrt{h}\ T(X)V, 
    \label{sub}
\end{equation}
where $\mu,\nu=0,\cdot,D-1$. 
Suppose that this corresponds to a (sub)critical theory, that is, $D+c_M\le26$, where $c_M$ is the 
central charge of $S_M$. 
The beta-functionals of this theory should be a linear combination of the equations of motion 
derived from (\ref{restricted}). 
One obtains 

\begin{eqnarray}
\beta_{\mu\nu}(g) 
&=& R_{\mu\nu}+2\nabla_\mu\nabla_\nu\Phi-\nabla_\mu T\nabla_\nu T+\nabla_\mu\Phi\nabla_\nu F(T)
   +\nabla_\nu\Phi\nabla_\mu F(T) \nonumber \\
& & -\frac12g_{\mu\nu}[ 2\nabla\Phi\cdot\nabla F(T)-\nabla^2F(T)], \\
      \label{betag}
\beta(\Phi) 
&=& \frac12V(T)-\frac12\nabla^2\Phi+(\nabla\Phi)^2-\frac{D-2}4\nabla\Phi\cdot\nabla F(T)
   +\frac{D-2}8\nabla^2F(T), \\
      \label{betaPhi}
\beta(T) 
&\propto& \nabla^2T-2\nabla\Phi\cdot\nabla T-V'(T)+f(T)[-\nabla^2\Phi+2(\nabla\Phi)^2], 
   \label{betaT}
\end{eqnarray}
where $F'(T)=f(T)$. 
$\beta_{\mu\nu}(g)$ and $\beta(\Phi)$ are determined so that they reduce to the well-know functionals 
when $T=0$. 
Note that $V(0)$ only appears in $\beta(\Phi)$, which is important below. 

Next, consider the following worldsheet theory, 
\begin{eqnarray}
S' 
&=& \frac1{4\pi}\int d^2\sigma\sqrt{h} \Bigl[ \nabla X^\mu\nabla X^\nu g_{\mu\nu}(X) 
    +\Phi(X)R_h \Bigr]+S_M(h)+\int d^2\sigma\sqrt{h}\ T(X)V \nonumber \\
& & +\frac1{4\pi}\int d^2\sigma\sqrt{h} \nabla Y^k\nabla Y^l \delta_{kl}, 
      \label{lifted}
\end{eqnarray}
where $k,l=1,\cdots,d$. 
The number $d$ is chosen so that $D+d+c_M(IR)>26$. 
The beta-functionals of $S'$ are just a sum of those of $S$ and those of free bosons $Y^k$. 
The conformal anomaly of $Y^k$ theory is simply  
\begin{equation}
T^{(Y)}{}^a{}_a = -\frac d{12}R_h. 
\end{equation}
Therefore, the beta-functionals of $S'$ have the same form as 
(\ref{betag})(\ref{betaPhi})(\ref{betaT}), with $V(T)$ replaced with $\tilde{V}(T)=V(T)+\frac d3$. 
Suppose that a solution exists for $\beta_{\mu\nu}(g)=0$ etc. for $S'$. 
At least when $f(T)=0$, a solution always exists as long as $V(T)>0$ in a range of $T$. 
Then it indicates that a background 
\begin{equation}
G_{MN} = \left[ 
\begin{array}{cc}
g_{\mu\nu}(X) & 0 \\ 0 & \delta_{kl}
   \label{supersoln}
\end{array}
\right], \hspace{5mm} \Phi=\Phi(X), \hspace{5mm} T=T(X), 
\end{equation}
is a classical solution of the effective action of the corresponding supercritical string. 

Consider in the opposite way. 
Suppose that there is an effective action of a supercritical string. 
The fields in the action are $G_{MN}(X,Y),\ \Phi(X,Y)$ and $T(X,Y)$. 
By making the ansatz (\ref{supersoln}), the equations of motion reduce to the conditions 
that (\ref{betag})(\ref{betaPhi})(\ref{betaT}), with $V(T)$ replaced with $\tilde{V}(T)$, all vanish. 
By the general covariance in $D+d$ dimensions, the equations of motion for 
$G_{MN}(X,Y),\ \Phi(X,Y)$ and $T(X,Y)$ should have the same form as 
those of $g_{\mu\nu}(X),\ \Phi(X)$ and $T(X)$, except for the potential. 
Therefore, the effective action should be 
\begin{equation}
S = \frac1{2\kappa^2}\int d^{D+d}x\sqrt{-G}\ e^{-2\Phi}\Bigl[ 
    R_G+4({\cal D}\Phi)^2+2f(T){\cal D}\Phi\cdot{\cal D} T-({\cal D} T)^2-2\tilde{V}(T) \Bigr], 
\end{equation}
where ${\cal D}$ and $R_G$ are the covariant derivative and the scalar curvature, respectively, 
defined by $G$. 

We have shown that the two string theories (\ref{sub}) and (\ref{lifted}) have essentially the same 
tachyon potential. 
Therefore, one can determine $V(T)$ by determining $\tilde{V}(T)$ in the way mentioned in the 
previous section. 

It should be emphasized that, when $V(T)$ is not always positive, 
the existence of a minimum of $V(T)$ does not always imply that there is 
a dynamical process to reach a static solution corresponding to the minimum. 
In this case, it is not yet certain that RG flows would be a nice tool to find the final state 
of a tachyon condensation.

\vspace{1cm}

\section{Discussion}

\vspace{5mm}

We have shown that RG flows can be use to describe on-shell tachyon condensations in supercritical 
strings. 
It seems natural to expect that the same strategy can apply to (sub)critical strings, and an 
observation was done in \cite{Suyama}. 
However, there is a difficulty in relating RG flows and time evolutions in these cases, as mentioned 
before. 
The main problem is that the Liouville field is spacelike. 
A naive Wick rotation would be possible to do for the case of 
a small perturbation around a critical string, 
but when there is a non-trivial dilaton background, such a Wick rotation would be impossible without 
changing the central charge. 
A recent study \cite{Schomerus} of the Liouville theory and its analytic continuation would be helpful 
to understand this situation. 
One might be possible to extract a spacetime physics from correlation functions of the Liouville 
theory obtained from the Euclidean theory via an analytic continuation. 

\vspace{5mm}

One may think that our procedure to analyze tachyon condensations can also apply to a condensation of 
bulk tachyons. 
We expect that it is possible in principle. 
To analyze a bulk tachyon condensation, one would consider a two-dimensional gravity perturbed by a 
term (\ref{perturb}) in which $V$ is the unit operator. 
Then, in the gauge-fixed action (\ref{fixedCFT}), the added term becomes the ordinary Liouville 
potential which is an exactly marginal perturbation. 
Therefore, the final state of the condensation is nothing but the flat background with infinitely 
large tachyon background. 
Recall that in the case of localized tachyons, in which $V$ is a non-trivial operator, the 
back-reaction to the background from the growing tachyon is implemented by a non-trivial RG flow, 
so that one does not need to deal with such a large tachyon background. 
In the bulk tachyon case, one has to treat such a background directly. 

When there is the Liouville potential in the time-direction, closed strings become heavier with time 
\cite{ST}. 
The same phenomenon occurs in the case of open string tachyons \cite{rolling1}\cite{rolling2}. 
In this case, since open strings become heavier due to a boundary interaction with a tachyonic 
background, the endpoints of the strings tend to pair-annihilate, and open strings turn into closed 
strings. 
This is a qualitative explanation of the decay of unstable D-branes. 
See also \cite{Martinec}. 
Is there a physical process which turns the heavy closed strings into the other lighter states? 
If not, the heavy closed strings will fill the space, resulting in a singularity like the one 
discussed in \cite{YZ}. 
Naively, there seems to be such a process; the reverse process of D-brane decay. 
Since the dilaton may become large without bound during a tachyon condensation, D-branes may 
become lighter. 
Then it is tempting to conjecture that closed strings would turn into a kind of D-branes, and the 
final state of the bulk tachyon condensation would be a theory whose fundamental degrees of freedom 
are D-branes. 

\vspace{5mm}

It would be worth mentioning on a classical solution of (\ref{simpleaction}) with $V_0=0,\ m^2>0$. 
In this case, $a_\pm$ in (\ref{a+-}) are purely imaginary, and the corresponding solution is periodic. 
However, one can see from the equations of motion that the exact solution cannot be periodic. 
In fact, the deviation from the periodic solution is the next-to-leading order. 
For example, $\dot{\phi}$ is the second order of fluctuations, and therefore the increase of $\phi$ 
is extremely small. 

How can one find this slowly varying behavior in the CFT side? 
To see this, one has to find a gauge-fixed worldsheet action without treating the tachyon vertex 
as a perturbation. 
$\alpha_\pm$ is purely imaginary when $c_m=25\ (V_0=0)$ since $Q=0$, but $Q=0$ is attained at 
$t\to+\infty$ limit. 
Since for large positive $t$ the background fields vary very slowly, one can approximate the 
background by a linear dilaton background with a small but non-zero $Q$. 
Then Re$(\alpha_\pm$) is non-zero, and the damping behavior still exists. 
The slow damping in the classical solution is due to the decrease of Re$(\alpha_\pm$) with time. 

It is very peculiar that no matter how large $m^2$ is, $T$ damps extremely slowly. 
Note that $\phi$ is almost zero, so the Einstein metric is equal to the string metric up to an overall 
factor, indicating that $T$ also damps slowly in the Einstein frame. 
It may be interesting if this phenomenon has some implications to cosmology. 
Note that the relation between the string metric $g^{(s)}_{\mu\nu}$ and the Einstein metric 
$g^{(E)}_{\mu\nu}$ is 
\begin{equation}
g^{(E)}_{\mu\nu}=e^{-\frac4{D-2}\Phi}g^{(s)}_{\mu\nu}. 
\end{equation}
Since we chose $\dot{\Phi}<0$, the Einstein metric is slowly expanding. 

The slow damping behavior is also plausible for the relation between RG flows and time evolutions. 
For $V_0=0$ case, the overall coefficient of the Liouville field in (\ref{fixedCFT}) vanishes. 
This is not a problem since by taking into account a change of central charge due to the perturbation, 
the coefficient, say $C$, does not vanish in a finite $\varphi_0$. 
However, $C$ should be decreasing as $\varphi_0\to+\infty$, so $C$ would be a function $C(\varphi_0)$. 
Then, to normalize the kinetic term of the Liouville theory, $\sqrt{C}$ should be included in 
$\varphi$, and as a result, the zero mode $\varphi_0'$ of the normalized field is 
\begin{equation}
\varphi_0' = \sqrt{C(\varphi_0)}\varphi_0. 
\end{equation}
If $C$ decreases too fast, the time $\varphi_0'$ does not correspond in one-to-one to the RG scale 
$\varphi_0$. 

\vspace{5mm}

One may be able to consider a rolling of $T$ in which initially $V(T)>0$ and finally $V(T)<0$. 
Assuming, for simplicity, that the simple action (\ref{simpleaction}) is valid in this case. 
In this case, $\phi=0$ does not imply $\dot{T}=0$ because of (\ref{eq}), and $\phi$ can be positive. 
Then $h=0$ solution is no longer a stable solution, as one can see from (\ref{stability}). 
It will be difficult to understand the late time behavior of this solution, but it would be 
anticipated that a simple linear dilaton solution cannot be the final configuration. 
This phenomenon would be related to a behavior of the RG flows of two-dimensional gravity, mentioned 
in \cite{Tseytlin}, which does not exist in the RG flows of two-dimensional field theory.

\vspace{2cm}

\begin{flushleft}
{\Large \bf Acknowledgments}
\end{flushleft}

\vspace{5mm}

I would like to thank T.Asakawa and S.Terashima for valuable discussions. 
I would also like to thank the Yukawa Institute for Theoretical Physics at Kyoto University. 
Discussions during the YITP workshop YITP-W-99-99 on "String Theory and Quantum Field Theory " were 
useful to complete this work. 
This work was supported in part by JSPS Research Fellowships for Young Scientists.

\appendix

\vspace{1cm}

\section{Derivation of Liouville action}   \label{derivation}

\vspace{5mm}

Consider a two-dimensional field theory coupled to gravity which is defined by the following 
partition function, 
\begin{equation}
Z_{gravity} 
= \int \frac{{\cal D}_gg{\cal D}_gX}{vol(\mbox{diff})}e^{-S(X;g)-\mu_0\int d^2\sigma\sqrt{g}}. 
\end{equation}
Suppose that $S(X;g)$ is conformally invariant and its trace anomaly does not have a constant term. 
Note that ${\cal D}_gX$, ${\cal D}_gg$ are defined in terms of $g$. 

One can fix the diffeomorphism invariance by choosing the following form of the metric 
\begin{equation}
g' = e^\varphi\hat{g}(\tau), 
\end{equation}
where $\hat{g}(\tau)$ is a representative metric parametrized by the moduli parameters $\{\tau\}$, 
in each orbit. 
Then the gauge-fixed form of the partition function is 
\begin{equation}
Z_{gravity} = \int d\tau{\cal D}_{g'}\varphi{\cal D}_{g'}b{\cal D}_{g'}c{\cal D}_{g'}X
              \ e^{-S(X;g')-S(b,c;g')-\mu_0\int d^2\sigma\sqrt{g'}}. 
\end{equation}

The measure ${\cal D}_{g'}\varphi$ is not a convenient one since its dependence on $\varphi$ itself 
is complicated. 
It is assumed in \cite{D}\cite{DK} 
that $Z_{gravity}$ is equivalent to the following partition function, 
\begin{equation}
Z = \int d\tau{\cal D}_{\hat{g}}\varphi{\cal D}_{\hat{g}}b{\cal D}_{\hat{g}}c{\cal D}_{\hat{g}}X
    \ e^{-S(X;\hat{g})-S(b,c;\hat{g})-S(\varphi;\hat{g})}, 
\end{equation}
where 
\begin{equation}
S(\varphi;\hat{g}) 
= \int d^2\sigma\sqrt{\hat{g}}\Bigl[ a(\hat{\nabla}\varphi)^2+b\hat{R}\varphi+\mu e^{c\varphi} \Bigr]. 
   \label{assumedaction}
\end{equation}
This can be rephrased as the equivalence of measures, 
\begin{equation}
{\cal D}_{g'}\varphi{\cal D}_{g'}b{\cal D}_{g'}c{\cal D}_{g'}X 
= {\cal D}_{\hat{g}}\varphi{\cal D}_{\hat{g}}b{\cal D}_{\hat{g}}c{\cal D}_{\hat{g}}X
  \ e^{-S(\varphi;\hat{g})}, 
\end{equation}
since $S(X;\hat{g}),\ S(b,c;\hat{g})$ are conformally invariant. 

The metric $g'$ can also be written as $g'=e^{\varphi'}e^\sigma\hat{g}$ for an arbitrary function 
$\sigma$. 
Then, in a similar way, one obtains 
\begin{equation}
{\cal D}_{g'}\varphi'{\cal D}_{g'}b{\cal D}_{g'}c{\cal D}_{g'}X 
= {\cal D}_{e^\sigma\hat{g}}\varphi'{\cal D}_{e^\sigma\hat{g}}b
  {\cal D}_{e^\sigma\hat{g}}c{\cal D}_{e^\sigma\hat{g}}X
  \ e^{-S(\varphi';e^\sigma\hat{g})}, 
\end{equation}
this implies 
\begin{equation}
{\cal D}_{\hat{g}}\varphi{\cal D}_{\hat{g}}b{\cal D}_{\hat{g}}c{\cal D}_{\hat{g}}X
\ e^{-S(\varphi;\hat{g})} 
= {\cal D}_{e^\sigma\hat{g}}\varphi'{\cal D}_{e^\sigma\hat{g}}b
  {\cal D}_{e^\sigma\hat{g}}c{\cal D}_{e^\sigma\hat{g}}X
  \ e^{-S(\varphi';e^\sigma\hat{g})}. 
\end{equation}
Therefore, one concludes that the total theory $Z$ must be conformally invariant. 
This requirement determines the coefficient $a,b,c$ in $S(\varphi;\hat{g})$. 

Let us determine $S(\varphi;\hat{g})$. 
For simplicity, suppose $\mu=0$. 
We assume that this is possible by tuning $\mu_0$ to the right value. 
One can show that 
\begin{equation}
{\cal D}_{e^\sigma\hat{g}}X{\cal D}_{e^\sigma\hat{g}}b{\cal D}_{e^\sigma\hat{g}}c
= e^{\frac{c_m-26}{48\pi}S_L(\sigma;\hat{g})}
  {\cal D}_{\hat{g}}X{\cal D}_{\hat{g}}b{\cal D}_{\hat{g}}c,
\end{equation}
where $c_m$ is the central charge of the matter part, and 
\begin{equation}
S_L(\sigma;\hat{g}) 
= \int d^2\sigma\sqrt{\hat{g}}\Bigl[ \frac12(\hat{\nabla}\sigma)^2+\hat{R}\sigma \Bigr]. 
\end{equation}
It should be noted that the Liouville potential in $S_L$ is absent since it has been assumed that 
the trace anomaly of the matter part does not have a constant term. 
See chapter 3 of \cite{Polchinski}. 

There is another argument for the absence of the Liouville potential. 
Suppose that the conformal anomaly only comes from the measure, 
\begin{equation}
{\cal D}_{e^\sigma\hat{g}}X = e^{\frac c{48\pi}S_L(\sigma;\hat{g})}{\cal D}_{\hat{g}}X. 
   \label{Weyl}
\end{equation}
Then the following equality
\begin{eqnarray}
{\cal D}_{e^{\sigma_1}e^{\sigma_2}\hat{g}}X 
&=& e^{\frac c{48\pi}S_L(\sigma_1+\sigma_2;\hat{g})}{\cal D}_{\hat{g}}X \nonumber \\
&=& e^{\frac c{48\pi}S_L(\sigma_1;e^{\sigma_2}\hat{g})}e^{\frac c{48\pi}S_L(\sigma_2;\hat{g})}
    {\cal D}_{\hat{g}}X. 
\end{eqnarray}
or 
\begin{equation}
S_L(\sigma_1+\sigma_2;\hat{g}) = S_L(\sigma_1;e^{\sigma_2}\hat{g})+S_L(\sigma_2;\hat{g}), 
\end{equation}
should hold. 
However, the general Liouville action  
\begin{equation}
S_L(\sigma;\hat{g},\mu') 
= \int d^2\sigma\sqrt{\hat{g}}\Bigl[ \frac12(\hat{\nabla}\sigma)^2+\hat{R}\sigma+\mu' e^{\sigma} 
  \Bigr]
\end{equation}
satisfies instead, 
\begin{equation}
S_L(\sigma_1+\sigma_2;\hat{g},\mu') 
= S_L(\sigma_1;e^{\sigma_2}\hat{g},\mu')+S_L(\sigma_2;\hat{g},\mu')
  -\mu'\int d^2\sigma\sqrt{\hat{g}}\ e^{\sigma_2}. 
\end{equation}
Therefore, the Weyl transformation (\ref{Weyl}) of the measure is consistent only if $\mu'=0$. 

Since $\mu=0$, the action of $\varphi$ is free, up to an unusual coupling to the worldsheet curvature. 
So it is natural to assume that ${\cal D}_{\hat{g}}\varphi$ is the measure of a free boson. 
Then its Weyl transformation is 
\begin{equation}
{\cal D}_{e^\sigma\hat{g}}\varphi = e^{\frac1{48\pi}S_L(\sigma;\hat{g})}{\cal D}_{\hat{g}}\varphi. 
\end{equation}
Therefore, to maintain the conformal invariance, $S(\varphi;\hat{g})$ must satisfies 
\begin{equation}
S(\varphi-\sigma;e^\sigma\hat{g}) = S(\varphi;\hat{g})+\frac{c_m-25}{48\pi}S_L(\sigma;\hat{g}). 
\end{equation}
This determines $2a=b=\frac{25-c_m}{48\pi}$, and therefore 
\begin{equation}
S(\varphi;\hat{g}) 
= \frac{25-c_m}{48\pi}\int d^2\sigma\Bigl[ \frac12(\hat{\nabla}\varphi)^2+\hat{R}\varphi \Bigr]. 
   \label{varphiaction}
\end{equation}

Next, let us consider the case $\mu\ne0$. 
As long as Re $c\ne0$ in (\ref{assumedaction}), 
$\varphi$ becomes free at $\varphi\to +\infty$ or $\varphi\to -\infty$. 
So around an either limit, $S(\varphi;\hat{g})$ should approach (\ref{varphiaction}). 
Therefore $a,b$ are the same values as above. 
Then, $c$ is determined by requiring that $e^{c\varphi}$ is a marginal operator of the CFT 
(\ref{varphiaction}). 
In the Liouville theory, the Liouville potential is actually an exactly marginal operator. 
So the value of $c$, determined by treating $e^{c\varphi}$ as a perturbation, also correct 
for a finite $\mu$.

\end{document}